\documentclass[10pt]{article}% scrbook, scrreprt, scrlettr, scrartcl
\usepackage[body={17cm, 23cm, centered}]{geometry}

\usepackage{tocloft}
\usepackage{subcaption}
\usepackage[utf8]{inputenc}
\usepackage{epsfig,verbatim,cite}
\usepackage{lmodern}
\usepackage{braket}
\usepackage[T1]{fontenc}
\usepackage{mathtext,marvosym,textcomp}
\usepackage{mathtools}
\usepackage{slashed}
\usepackage[usenames,dvipsnames,svgnames,table]{xcolor}
\usepackage{tikz}
\usepackage{tikz-cd}
\usepackage[ normalem]{ulem}

\usepackage{float}

\numberwithin{equation}{section}

\usepackage{epsfig,amsmath,amsfonts,amssymb,arydshln}
\usepackage[makeroom]{cancel}
\usepackage{multirow}

 \usepackage[debug,pageanchor=false]{hyperref}
\hypersetup{colorlinks=true,linktocpage,breaklinks,
            urlcolor=blue,
            linkcolor=red,
            citecolor=blue
            }

\def\ph{\phantom}

\newcommand\bqa {\begin{eqnarray}}
\newcommand\eqa {\end{eqnarray}}

\newcommand{\bear}{\begin{array}}
\newcommand{\enar}{\end{array}}

\newcommand{\be}{\begin{equation}}
\newcommand{\ee}{\end{equation}}
\newcommand{\bea}{\begin{eqnarray}}
\newcommand{\eea}{\end{eqnarray}}

\begin{document}

\renewcommand{\contentsname}{}
\renewcommand{\refname}{\begin{center}References\end{center}}
\renewcommand{\abstractname}{\begin{center}\footnotesize{\bf Abstract}\end{center}} 

\begin{titlepage}
\ph{preprint}

\vfill

\begin{center}
   \baselineskip=16pt
   {\large \bf On the late time behavior of thermal two point function  in curved space-time
   }
   \vskip 2cm
    Dmitrii V. Diakonov$^{a,b}$\footnote{\tt dmitrii.dyakonov@phystech.edu}, Gleb S. Zverev$^{a}$\footnote{\tt zverev.gs@phystech.edu}
       \vskip .6cm
             \begin{small}
                          {\it
                          $^a$Moscow Institute of Physics and Technology, 
                          Laboratory of High Energy Physics\\
                          Institutskii per., 9, 141702, Dolgoprudny, Russia\\
                          $^b$Institute for Information Transmission Problems, Bol’shoi Karetnyi per., 19,  127994, Moscow, Russia
                          } \\ 
\end{small}
\end{center}

\vfill 
\begin{center} 
\textbf{Abstract}
\end{center} 
\begin{quote}
We investigate the late-time behavior of the thermal two-point function for a Bose gas in a curved space-time with a Killing horizon. We demonstrate that the late-time behavior undergoes a sharp transition at a critical temperature, reminiscent of a phase transition. Contrary to expectations for typical gases, the relaxation time does not vanish at high temperatures but instead saturates to a constant value that depends on the imaginary part of the lowest quasinormal modes.
\end{quote}

\vfill
\setcounter{footnote}{0}
\end{titlepage}

%\tableofcontents
%\setcounter{page}{1}

\section{Introduction}

In a large number of physical systems, the correlation function decays exponentially in time:
\begin{align}
    \langle O(t) O(0)\rangle \sim e^{-\frac{t}{\tau}},
\end{align}
where $\tau$ is a relaxation time; see \cite{Lucas:2018wsc} and references therein. In thermal systems, $\tau$ is a function of temperature:
\begin{align}
    \tau=\tau(T). 
\end{align}
As a simple example, the velocity correlation function of the Brownian motion of spherical particles is given by \cite{bird2002transport}:  
\begin{align}
    \langle v(t) v(0)\rangle \sim e^{-\frac{t}{\tau}}, 
\end{align}
where the relaxation time: 
\begin{align}
    \tau(T)= \frac{m}{6\pi R \eta(T)},
\end{align}
depends on the mass $(m)$ and radius $(R)$ of the particle and the viscosity $(\eta)$ of a gas or a fluid, which depends on its temperature. For example, for classical gas
, viscosity increases proportionally with the square root of the temperature\cite{bird2002transport}: 
\begin{align}
    \eta(T)\sim \sqrt{T}.
\end{align}
Hence, for classical gas, the relaxation time decreases with increasing temperature, which is a common situation for gases. In contrast, in liquids, viscosity usually decreases with increasing temperature. However, near the glass transition, the viscosity of liquids can increase exponentially \cite{bird2002transport}: 
\begin{align}
    \eta(T)\sim e^{\frac{A}{T-T_0}}.
\end{align}
Note that the glass transition is not considered as a phase transition; rather, it is a phenomenon where the viscosity of certain liquids can increase by up to 13 orders of magnitude. This means that the relaxation time almost vanishes in such a limit. Hence, all these points show that the late-time correlation function reflects how fast the system "forgets" its initial state---i.e., how quickly it returns to equilibrium. Moreover, if the correlation time changes significantly at some temperature, it reflects a change in the physical properties of the gas or liquid at that temperature. 
Note also that for all examples discussed above the exponential decay arise due to interactions. Nevertheless, similar exponential decay appears in free field theory.

In four-dimensional Minkowski space-time, in the high-temperature limit or in the massless case, the late-time behavior of the thermal two-point function at coincident spatial points is defined as follows:
\begin{align}
    W_\beta(t)\sim \sum_{n=-\infty}^\infty \frac{1}{(t+i\beta n)^2} \sim T^2 e^{-\frac{2\pi}{\beta} t}, \quad \beta=\frac{1}{T}.
\end{align}
Hence, the relaxation time is given by: 
\begin{align}
    \tau(T) =\frac{1}{2\pi T},
\end{align}
which decreases with increasing temperature, a common situation for gases. A similar exponential decay occurs for spin $1/2$, $1$, and $3/2$ thermal two-point functions \cite{Weldon:2000pe}. Moreover, such a behavior is present in thermal CFT. Indeed, the two-dimensional CFT thermal two-point function can be completely fixed using conformal invariance \cite{Skenderis:2008dg,Akhmedov:2022gzt,Sadekov:2021rpy}: 
\begin{align}
\label{CFT}
   \langle O(t,x) O(0)\rangle_\beta= \left(\frac{2\pi}{\beta}\right)^{2\Delta}\left(2 \cosh\left(\frac{2\pi x}{\beta}\right)-2 \cosh\left(\frac{2\pi t}{\beta}\right) \right)^{-\Delta},
\end{align}
where $\Delta$ is the conformal dimension. Hence, at late time the thermal two-point function decays exponentially in time: 
\begin{align}
    \langle O(t) O(0)\rangle \sim e^{-\frac{2\pi }{\beta}\Delta t}.
\end{align}
In this case, the relaxation time depends on the conformal dimension and also decreases with increasing temperature:
\begin{align}
    \tau(T)=\frac{1}{2\pi T \Delta} .
\end{align}
Such a decay is not a dynamical instability. This decay simply tells us that at finite temperature, correlations are lost over a characteristic timescale $\tau_{\text{T}} \sim 1/T$, which follows naively from an intuitive argument based on the Heisenberg energy-time uncertainty principle: $\Delta E \Delta t \gtrsim 1$ with $\Delta t=\tau$ and $\Delta E=T$. It follows that the relaxation time has a bound \cite{Lucas:2018wsc}: 
\begin{align}
\label{tau_T}
    \tau \gtrsim \tau_T= \frac{1}{T}. 
\end{align}

In a space-time with a Killing horizon, one can define quasinormal modes that are purely ingoing near the horizon. These quasinormal modes describe the decay of perturbations and have complex frequencies such that the field decays exponentially at late times:
\begin{align}
    \phi(t) \sim e^{-t  \omega_I},
\end{align}
where $\omega_I$ is the imaginary part of the lowest quasinormal frequency. Hence, one can expect that for such a system the relaxation time is given by: 
\begin{align}
\label{tau_Q}
    \tau_Q = \frac{1}{\omega_I}.
\end{align}
Thus, for systems with a Killing horizon, we have two possibilities for choosing the characteristic relaxation time: the first is due to thermal fluctuations \eqref{tau_T} and the second is due to the presence of quasinormal modes \eqref{tau_Q}.

In this note, we show with a few examples that the late-time behavior of the thermal two-point function of a Bose gas in a curved space-time with a Killing horizon can change at a certain temperature $T_c$: 
\begin{equation}
    \begin{gathered}
        W_{\beta}(t \to +\infty) \sim 
        \left\{\begin{matrix}
        \hspace{0.2em}e^{- 2\pi T t},\hspace{1em} T <T_c
         \\
        \hspace{0.2em}e^{-2 \pi T_c t},\hspace{1em} T >T_c
        \end{matrix}\right.  
    \end{gathered}.
\end{equation}
In the first case, the relaxation time saturates the thermal bound $\tau\sim \tau_T$, and in the second case, the relaxation time is determined by the imaginary part of the quasinormal modes $\tau\sim \tau_Q$.

\section{Space-times with Killing horizons}
Due to Hawking-type radiation \cite{Hawking:1975vcx}, space-times with Killing horizons are usually endowed with a natural (canonical) temperature, which depends on the geometry of the space-time, i.e., a temperature proportional to the surface gravity:
\begin{align}
    \beta_c=\frac{2\pi}{\kappa}. 
\end{align}
Space-times with Killing horizons admit a Euclidean version by analytic continuation $t\to i\tau$. The requirement of regularity of the Euclidean metric, i.e., the absence of a conical singularity, imposes that $\tau$ is compact with period $\beta_c$. Hence, it is natural to consider a thermal gas with the Planckian density matrix at the canonical temperature. Nevertheless, in space-times with Killing horizons, one can also consider a thermal gas of exact mode with the Planckian density matrix at an arbitrary temperature, different from the canonical one. However, if the temperature is different from the canonical one, then the thermal two-point function does not possess Hadamard properties on the Killing horizons and the back-reaction on the background geometry is strong (see, for example: \cite{Candelas:1980zt, Fulling:1977zs,Akhmedov:2019esv, Akhmedov:2020ryq,Akhmedov:2020qxd, Anempodistov:2020oki, Akhmedov:2021cwh, Bazarov:2021rrb,Akhmedov:2022qpu}). Moreover, if the temperature of the system is less than the Hawking one, then the theory becomes unstable due to the presence of a tachyon excitation \cite{Diakonov:2023hzg}. Note also that if we consider the Euclidean functional integral on a conical manifold with a generic period in imaginary time ($\beta$ -- the inverse temperature), then the leading divergent contribution to the free energy depends on $\beta$ \cite{Kabat:1995eq}. Hence, for arbitrary $\beta$, it is not clear how to renormalize the divergent contribution, since it depends on the state of the system. Nevertheless, if the temperature is equal to the canonical one, which can be considered as a geometrical characteristic of the space, then one can subtract this divergence by renormalizing curvature couplings in the gravitational action \cite{Solodukhin:1994yz, Fursaev:1994ea, Solodukhin:1995ak,Diakonov:2025wtt}. 

The other difference between space-times with Killing horizons and without them is that the single-particle energy spectrum of the theory does not depend on the mass. Indeed, the single-particle states can be found using the Klein-Gordon equation for the modes $\psi_\omega(x) =e^{-i \omega t} f_{\omega}(x)$:
\begin{align}
   \left(-\Box+m^2+\xi R\right) \psi_\omega(x)=0 \quad \to \quad \left(-\omega^2 g^{00}-\triangle_3+m^2 +\xi R\right) f_\omega(x)=0.
\end{align}
From the fact that $g^{00}\omega^2$ diverges near the Killing horizon since $g^{00}$ goes to infinity, the mass term is negligibly small and the spectrum of the theory is defined as $\omega \in (0,\infty)$. This property will be important to find the late-time behavior of the thermal two-point function since in such a case we will obtain that: 
\begin{align}
\label{prodformula}
    W_\beta(t)\sim \int_{-\infty}^\infty d \omega \ e^{i\omega t } \frac{\sinh(\pi \omega)}{e^{\beta \omega}-1} \prod_{\omega_i} \frac{1}{\omega-\omega_i},
\end{align}
where the integral over $\omega$ runs from $-\infty$ to $\infty$ and $\omega_i$ are the poles that are related to the quasinormal modes. Due to such an expansion, one can find the late-time asymptotic behavior of the thermal two-point function using the residue theorem and show that the leading contributions come from the poles which are closer to the real axis. This is not a general expansion of the thermal two-point function in a space-time with a Killing horizon, but in the cases that we will consider, it is. Note also that a similar product formula was obtained in \cite{Dodelson:2023vrw} for the holographic thermal two-point function in AdS/CFT and in \cite{Grewal:2024emf,Akhmedov:2020qxd} for de Sitter space-time.

\subsection{de Sitter}
The metric of the de Sitter static patch is given by:
\begin{align}
\label{metricdS}
ds^2=-\left(1-r^2\right) dt^2+\left(1-r^2\right) ^{-1} dr^2+r^2 d\Omega_{d-2}^2,
\end{align}
where we set the de Sitter radius to $R = 1$. One of the main properties of this metric is the existence of a time-like Killing vector, which allows one to introduce the notion of energy and define a thermal state with a Planckian distribution for the exact modes at an inverse temperature $\beta$. The static patch is bordered by a Killing horizon at $r=1$, where the metric degenerates.

For the case when $d>2$, the mode expansion of the scalar field operator that obeys the canonical commutation relations is defined as follows \cite{Akhmedov:2021cwh}:
\begin{align}
\phi(t,r,\Omega)=\int _0^\infty d\omega\sum_{jk} \left(\phi_{\omega j k} a_{\omega j k}+\phi^*_{\omega j k} a^\dagger_{\omega j k}\right),
\label{harmexp}
\end{align}
where the creation and annihilation operators obey standard commutation relations. The mode function $\phi_{\omega j k}$ is a solution of the Klein–Gordon equation and is given by:
\begin{align}
\phi_{\omega j k}(t,r,\Omega)=\frac{1}{\sqrt{2 \omega}}e^{-i \omega  t} R_{\omega j}(r) Y_{j k}(\Omega), \label{phiRY}
\end{align}
where $Y_{j k}(\Omega)$ are hyperspherical harmonics and 
\begin{gather}
\label{Rmode}
R_{\omega j}(r)=\frac{ \Gamma \left(\frac{d+2 j-2 i \nu -2 i \omega-1}{4} \right) \Gamma \left(\frac{d+2 j+2 i \nu -2 i \omega-1}{4}\right) }{\sqrt{2 \pi}  \Gamma \left( i \omega\right)\, \Gamma\left(\frac{d-1}{2}+j\right)} \times \\ \nonumber  \times r^j (1-r^2)^{\frac{i\omega }{2}}      {}_2 F_1\left(\frac{ i \omega - i \nu+ j+\frac{d-1}{2}}{2},\frac{ i \omega + i \nu+ j+\frac{d-1}{2}}{2}, j+\frac{d-1}{2},r^2\right),
\end{gather}
where $_2F_1 $ is the hypergeometric function and
\begin{equation}
    \nu =\sqrt{m^2-\left(\frac{d-1}{2}\right)^2}.
    \label{nunu}
\end{equation}

At coincident spatial points $r_1=r_2= 0$ and $\Omega_1=\Omega_2=\Omega$, the thermal two-point function with an arbitrary temperature (not equal to the Gibbons-Hawking temperature $\beta=2\pi$) is given by \cite{Akhmedov:2021cwh,Grewal:2024emf}: 
\begin{gather}
\nonumber
  W_\beta(t=t_1-t_2)= \sum_{jk} Y_{j k}(\Omega) Y_{j k}(\Omega)  \int_{0}^{\infty}d\omega    \frac{ \left|\Gamma \left(\frac{d+2j-2 i \nu -2 i \omega-1}{4} \right) \Gamma \left(\frac{d+2j+2 i \nu -2 i \omega-1}{4}\right) \right|^2}{4 \pi \omega |\Gamma \left( i \omega\right)| ^2\, \Gamma^2\left(\frac{d-1}{2}+j\right)}\times \\ \times\Bigg[
e^{i\omega t} \frac{1}{e^{\beta \omega }-1}
+
e^{-i\omega t}  \left(1+\frac{1}{e^{\beta \omega }-1}\right)\Bigg].
\label{twopoinds2}
 \end{gather}
Here the main point is that since the spectrum of the theory starts from zero, we can rewrite \eqref{twopoinds2} as an integral over $\omega$ from $-\infty$ to $+\infty$:
\begin{gather}
\nonumber
  W_\beta(t)= \\=\sum_{jk} Y_{j k}(\Omega) Y_{j k}(\Omega)  \int_{-\infty}^{\infty}d\omega    \frac{ \left|\Gamma \left(\frac{d+2j-2 i \nu -2 i \omega-1}{4} \right) \Gamma \left(\frac{d+2j+2 i \nu -2 i \omega-1}{4}\right) \right|^2}{4 \pi \omega |\Gamma \left( i \omega\right)| ^2\, \Gamma^2\left(\frac{d-1}{2}+j\right)}
 \frac{e^{i\omega t}}{e^{\beta \omega }-1}
 \label{wpoles}
.
 \end{gather}
The integrand in \eqref{wpoles} has poles at:
\begin{align}
\omega=\pm i \left(\frac{d-1}{2}+j+2n\right)\pm  \nu, \quad n\in \mathbf{Z^+}, \quad \omega = \frac{2 \pi i k}{\beta},\quad k\in \mathbf{Z}, \ k\neq 0.
\end{align}
The first set corresponds to the familiar frequencies of the quasinormal modes in de Sitter space-time \cite{Lopez-Ortega:2006aal}, and the second set corresponds to the Matsubara frequencies. 
In the limit $t\rightarrow \infty$, the leading contributions come from the poles which are closest to the real axis; hence, all terms with $j>0$ are suppressed. Therefore, in this limit: 
\begin{gather}
  \label{limtwopoin}
  W_\beta(t)
  \sim
\nonumber
\int_{-\infty}^{\infty}d\omega\frac{ e^{i\omega t}}{4 \pi \Gamma^2 \left(\frac{d-1}{2}\right) } \frac{\sinh(\pi \omega)}{e^{\beta \omega }-1} \Bigg|\Gamma \left(\frac{1}{4} (d-1-2 i \nu -2 i \omega)\right) \Gamma
   \left(\frac{1}{4} (d-1+2 i \nu -2 i \omega)\right) \Bigg|^2. 
 \end{gather}
 Using the product formula for the gamma function:
 \begin{align}
 \label{gamma}
     \displaystyle \Gamma (z)={\frac {e^{-\gamma z}}{z}}\prod _{n=1}^{\infty }\left(1+{\frac {z}{n}}\right)^{-1}e^{z/n},
 \end{align}
the thermal two-point function can be written in the form \eqref{prodformula}. Using the residue theorem, the late-time asymptotic behavior can be easily obtained: 
\begin{equation}
\label{beh}
W_\beta(t\to \infty) \approx
 \begin{cases}
  e^{-t \frac{d-1}{2}} \left(C_+ e^{i\nu t} + C_- e^{- i\nu t}\right) &, \text{if} \ \  \beta < \frac{4\pi}{d-1} \\
   C_\beta e^{-t \frac{2\pi}{\beta}} &, \text{if} \ \ \beta> \frac{4\pi}{d-1}
   \end{cases}.
\end{equation}
The asymptotic behavior of the propagator changes at $\beta=\frac{4\pi}{d-1}$. As $\beta$ increases, the poles of the second set move towards the real axis. When $\beta>\frac{4\pi}{d-1}$, the poles of the second set dominate and the large $t$ behavior of the propagator changes accordingly. Note also that for $\beta=2 \pi\frac{k}{n}$, there are partial cancellations of poles at Matsubara frequencies since: 
\begin{align}
    \sinh\left(\pi  \times \frac{2 \pi i k}{\beta}\right)=0.
\end{align}
Thus, for such temperatures and with the condition $\beta>\frac{4\pi}{d-1}$, the closest poles may be the quasinormal modes poles, and the late-time behavior is the same as for $\beta<\frac{4\pi}{d-1}$. As a simple example, one can consider the canonical temperature $\beta=2\pi$; then all Matsubara poles are canceled and the late-time behavior is defined by the quasinormal modes poles closest to the real axis. Note also that the late-time behavior is sensitive to the partial cancellation of poles only for low temperatures, since for high temperatures the quasinormal modes poles dominate. 

As a result, we can see that at low temperatures the relaxation time decreases with increasing temperature: $\tau\sim \frac{1}{T}=\tau_T$, as in flat space-time. However, at high temperatures the relaxation time is determined by the imaginary part of the quasi-normal modes $\tau\sim \frac{1}{\omega_I}=\tau_Q$, and does not decrease to zero as one might expect.

The question immediately arises: do these properties manifest themselves in the partition function and, consequently, in thermodynamic quantities such as entropy, heat capacity, etc.?

The free energy is given by \cite{Anninos:2020hfj,Akhmedov:2021cwh}:
\begin{align}
\label{FD2}
 F_\beta=-\frac{1}{2^{d-1}\pi} \int_\gamma dy\frac{\frac{\pi  y }{\beta }\coth \left(\frac{\pi  y}{\beta }\right)-1}{2 y^2}\frac{e^{i \nu y}}{\sinh ^{d-1}\left(\frac{|y|}{2}\right)}.
\end{align}
The main contribution to the free energy in the IR limit (large mass) is defined as follows \cite{Akhmedov:2021cwh}:
\begin{equation}
F_\beta\approx
 \begin{cases}
-\frac{ 1}{ 2^{d-1}\beta\left[i   \sin \left(\frac{ \beta }{2}\right)\right]^{d-1}}  e^{- \beta  m  } &,
 \text{if} \ \  \beta <  2 \pi \\ -\frac{(i m)^{d-2}}{  2^{d+1} \pi^2 i (d-2)!}
 \left[\frac{  2 \pi^2   }{\beta }\cot \left(\frac{2  \pi^2  }{\beta }\right)-1\right]  e^{- 2 \pi m}
 &, \text{if} \ \ \beta> 2 \pi
   \end{cases}.
\end{equation}
Thus, in the high-temperature limit, one obtains $ F_\beta \sim e^{- \beta  m  } $, while in the low-temperature limit, $ F_\beta \sim e^{- 2 \pi  m  } $. That is, the behavior of the leading contribution to the free energy changes at $\beta = 2 \pi$, not at $\beta=\frac{4\pi}{d-1}$. Hence, no phase transition occurs at $\beta=\frac{4\pi}{d-1}$.

\subsection{Planar BTZ black hole}
As another example, we consider a non-rotating BTZ black hole. The metric is given by: 
\begin{equation}
    \begin{gathered}
        ds^{2} = -\left(1 - \frac{r_{+}^{2}}{r^{2}}\right)\frac{r^{2}dt^{2}}{R^{2}} + \frac{R^{2}dr^{2}}{r^{2}\left(1 - \frac{r_{+}^{2}}{r^{2}}\right)} + r^{2}d\varphi^{2},
        \\
        r_{+}^{2} = MR^{2},
    \end{gathered}
\end{equation}
where $r_{+}$ is the radius of the black hole, $R$ is the radius of $\mathrm{AdS}_{3}$, and $M$ is the mass of the black hole. The Killing horizon corresponds to $r = r_{+}$, where the metric degenerates. Throughout the following, we set $R = 1$ and $M = 1$ for convenience.

The mode expansion of the scalar field operator that obeys the canonical commutation relations is defined as follows:
\begin{align}
\phi(t,r,\Omega)=\int _0^\infty d\omega\sum_{n} \left(\phi_{\omega n} a_{\omega n}+\phi^*_{\omega n} a^\dagger_{\omega n}\right),
\end{align}
where the creation and annihilation operators obey the canonical commutation relations. The mode function $\phi_{\omega n}$ is a solution of the Klein–Gordon equation and is given by \cite{Papadodimas:2012aq}:
\begin{align}
\phi_{\omega n}(t,r,\Omega)=\frac{1}{\sqrt{2 \omega}}e^{-i \omega  t} e^{i n \phi} R_{\omega n}(r),
\end{align}
where: 
\begin{gather}
        R_{\omega n} = \frac{\Gamma(in)}{\sqrt{2}\Gamma(i\omega)}e^{\frac{\pi\omega}{2}}\left(r^{2}-1\right)^{-\frac{i\omega}{2}}\Bigg(e^{\frac{\pi n}{2}}r^{-in}{}_2 F_1\bigg[\frac{2 - \Delta - i(n+\omega)}{2},\frac{ \Delta - i(n+\omega)}{2},1-in,r^{2}\bigg] 
        + \\
        + \nonumber
        e^{-\frac{\pi n}{2}}r^{i n}\frac{\Gamma(1-i n)\Gamma\big(\frac{\Delta +i(n-\omega)}{2}\big)\Gamma\big(\frac{\Delta +i(n+\omega)}{2}\big)}{\Gamma(1+in)\Gamma\big(\frac{\Delta -i(n-\omega)}{2}\big)\Gamma\big(\frac{\Delta -i(n+\omega)}{2}\big)}{}_2 F_1\bigg[\frac{2 - \Delta + i(n-\omega)}{2},\frac{\Delta+ i(n-\omega)}{2},1+in,r^{2}\bigg]\Bigg),
\end{gather}
and $\Delta = 1 + \sqrt{m^{2}+1}$. The thermal two-point function at coincident points on the conformal boundary $r_{1} = r_{2} = r \to +\infty$, $\varphi_{1} = \varphi_{2}$, has the form:
\begin{equation}
    W_{\beta}\left(t,r \to \infty\right) \approx \frac{r^{-2\Delta}}{4 \pi^{2}  \left|\Gamma (\Delta )\right|^{2}}\sum_{n}\int_{-\infty}^{+\infty}d \omega e^{i\omega t} \frac{\sinh (\pi  \omega )}{e^{\beta  \omega }-1}
    \left|\Gamma \left(\frac{\Delta+i (n-\omega )}{2}\right)\Gamma \left(\frac{\Delta+ i(n+\omega )}{2}\right)\right|^{2}.
\end{equation}
To find the late-time behavior, one should take into account infinitely many poles, since poles coming from gamma functions with different $n$ have the same imaginary parts. One should first take the sum over $n$ and then study the late-time behavior. The story becomes simpler in the case of a planar BTZ black hole. In this case, we can obtain a closed-form answer. By replacing the summation over $n$ with an integral over $k$, we can use a table integral \cite{gradshteyn2007}: 
\begin{gather}
    \int_{-i \infty}^{i \infty} ds \ \Gamma\left(\alpha +s\right)\Gamma\left(\beta +s\right)\Gamma\left(\gamma -s\right)\Gamma\left(\delta -s\right)=\\= \nonumber 2\pi i  \frac{\Gamma\left(\alpha+\gamma\right)\Gamma\left(\alpha+\delta\right)\Gamma\left(\beta+\gamma\right)\Gamma\left(\beta+\delta\right)}{\Gamma\left(\alpha+\beta+\gamma+\delta\right)},
\end{gather}
to arrive at the following expression:
\begin{gather}
\label{boundtwopoint}
        W_{\beta}\left(t,r\to \infty\right) \approx \frac{r^{-2\Delta}}{\Gamma(2\Delta)}\int_{-\infty}^{+\infty}\frac{d\omega}{2\pi}\hspace{0.2em}e^{i\omega t}\frac{\sinh{(\pi\omega)}}{e^{\beta\omega}-1}\Gamma(\Delta+i\omega)\Gamma(\Delta-i\omega).
    \end{gather}
Using the product formula for the gamma function \eqref{gamma}, the thermal two-point function can be written in the form \eqref{prodformula}. 

Note that for $\beta=2\pi$, we can recover the two-dimensional CFT thermal two-point function \eqref{CFT} with an inverse temperature that depends on the radius of the AdS space-time: $\beta=\frac{2\pi}{R}$ (we fixed $R=1$ above). It is not clear to us whether this boundary two-point function \eqref{boundtwopoint} for an arbitrary temperature $\beta\ne 2\pi$ corresponds to the two-point function in some two-dimensional theory. It is only clear that such a two-point function depends on three parameters $(\beta,R,\Delta)$.  
 
As in the previous example, we have two sets of poles: the Matsubara frequencies and those proportional to the quasinormal modes in BTZ \cite{Birmingham:2001hc}: 
\begin{gather*}
    \omega_{k} = \frac{2\pi i}{\beta}k,\hspace{0.5em} k\in\mathbb{Z}\setminus \{0\};\hspace{1em} \omega_{n} = \pm i\left(n+\Delta\right),\hspace{0.5em} n \in \mathbb{N}\cup\{0\}.
\end{gather*}
In the limit $t \to +\infty$, the leading contributions come from the poles that are closest to the real axis. Therefore, the asymptotic behavior of the propagator changes at $\beta=\frac{2\pi}{\Delta}$: 
\begin{equation}
    \begin{gathered}
        W_{\beta}(t \to \infty,r \to \infty) \simeq r^{-2\Delta}
        \left\{\begin{matrix}
        C_{\beta}\hspace{0.2em}e^{-\frac{2\pi}{\beta}t},\hspace{1em} \beta >\frac{2\pi}{\Delta}
         \\
        C_{\Delta}\hspace{0.2em}e^{-\Delta t},\hspace{1em} \beta <\frac{2\pi}{\Delta}
        \end{matrix}\right..
    \end{gathered}
\end{equation}
Note also that for $\beta=2 \pi\frac{k}{n}$, there is a partial cancellation of poles at Matsubara frequencies; for example, for $\beta=2\pi$, all Matsubara poles are canceled and the late-time behavior is defined only by the quasinormal modes poles closest to the real axis.

\section{Conclusion}
In this paper, we show with a few examples that the late-time behavior of the thermal two-point function of a Bose gas in a curved space-time with a Killing horizon can change at a certain temperature $(T_c)$: 
\begin{equation}
    \begin{gathered}
        W_\beta(t \to \infty) \sim
        \left\{\begin{matrix}
        \hspace{0.2em}e^{- 2\pi T t},\hspace{1em} T <T_c
         \\
        \hspace{0.2em}e^{-2 \pi T_c t},\hspace{1em} T >T_c
        \end{matrix}\right.
    \end{gathered}.
\end{equation}
In the first case, the relaxation time saturates the thermal bound $\tau\sim \tau_T$, and in the second case, the relaxation time is determined by the imaginary part of the quasi-normal modes $\tau\sim \tau_Q$.
Moreover, such behavior shows that: 
\begin{itemize}
    \item At temperature $T=T_c$, the relaxation time changes, which resembles a phase transition.
    \item At high temperatures, the relaxation time does not decrease to zero as one might expect for gases. 
\end{itemize} 
Note also that the inverse relaxation time at infinite temperature $1/\tau \sim T_c$ is sometimes referred to as the "tOmperature", since the thermal two-point function at infinite temperature has a thermal form, with the effective temperature $T_c$ \cite{Lin:2022nss,Milekhin:2024vbb}.  

We do not have yet a strong criterion to specify for which space-times the thermal two-point function exhibits such a behavior. However, we have verified it directly for the static de Sitter patch and the planar BTZ black hole. In all such cases, the key properties are that the space-time is curved, there is a horizon, and there exist quasinormal modes of the scalar field in such a background.

\section*{Acknowledgments}
The authors would like to thank Kirill Bazarov, Damir Sadekov, Edvard Musaev and Dmitry Ageev for helpful discussions. would like to thank E.T.Akhmedov for valuable discussions, sharing his ideas and correcting the text. The work of DDV was supported by the grant from the Foundation for the Advancement of Theoretical Physics and Mathematics ``BASIS'', and the state assignment of the Institute for Information Transmission Problems of RAS.

\bibliography{bib.bib}

\providecommand{\href}[2]{#2}\begingroup\raggedright\begin{thebibliography}{10}

\bibitem{Lucas:2018wsc}
A.~Lucas, ``{Operator size at finite temperature and Planckian bounds on quantum dynamics},'' \href{http://dx.doi.org/10.1103/PhysRevLett.122.216601}{{\em Phys. Rev. Lett.} {\bfseries 122} no.~21, (2019) 216601}, \href{http://arxiv.org/abs/1809.07769}{{\ttfamily arXiv:1809.07769 [cond-mat.str-el]}}.

\bibitem{bird2002transport}
R.~Bird, W.~Stewart, and E.~Lightfoot, {\em Transport Phenomena}.
\newblock J. Wiley, 2002.
\newblock \url{https://books.google.ru/books?id=wYnRQwAACAAJ}.

\bibitem{Weldon:2000pe}
H.~A. Weldon, ``{Thermal Green functions in coordinate space for massless particles of any spin},'' \href{http://dx.doi.org/10.1103/PhysRevD.62.056010}{{\em Phys. Rev. D} {\bfseries 62} (2000) 056010}, \href{http://arxiv.org/abs/hep-ph/0007138}{{\ttfamily arXiv:hep-ph/0007138}}.

\bibitem{Skenderis:2008dg}
K.~Skenderis and B.~C. van Rees, ``{Real-time gauge/gravity duality: Prescription, Renormalization and Examples},'' \href{http://dx.doi.org/10.1088/1126-6708/2009/05/085}{{\em JHEP} {\bfseries 05} (2009) 085}, \href{http://arxiv.org/abs/0812.2909}{{\ttfamily arXiv:0812.2909 [hep-th]}}.

\bibitem{Akhmedov:2022gzt}
E.~Akhmedov, H.~Epstein, and U.~Moschella, ``{The massless thermal field and the thermal fermion bosonization in two dimensions},'' \href{http://dx.doi.org/10.1007/JHEP09(2022)123}{{\em JHEP} {\bfseries 09} (2022) 123}, \href{http://arxiv.org/abs/2203.02747}{{\ttfamily arXiv:2203.02747 [hep-th]}}.

\bibitem{Sadekov:2021rpy}
D.~Sadekov, ``{Generalization of 2D gravity with the simplest noninvariant states},'' \href{http://dx.doi.org/10.1103/PhysRevD.105.125003}{{\em Phys. Rev. D} {\bfseries 105} no.~12, (2022) 125003}, \href{http://arxiv.org/abs/2111.05984}{{\ttfamily arXiv:2111.05984 [hep-th]}}.

\bibitem{Hawking:1975vcx}
S.~W. Hawking, ``{Particle Creation by Black Holes},'' \href{http://dx.doi.org/10.1007/BF02345020}{{\em Commun. Math. Phys.} {\bfseries 43} (1975) 199--220}. [Erratum: Commun.Math.Phys. 46, 206 (1976)].

\bibitem{Candelas:1980zt}
P.~Candelas, ``{Vacuum Polarization in Schwarzschild Space-Time},'' \href{http://dx.doi.org/10.1103/PhysRevD.21.2185}{{\em Phys. Rev. D} {\bfseries 21} (1980) 2185--2202}.

\bibitem{Fulling:1977zs}
S.~A. Fulling, ``{Alternative Vacuum States in Static Space-Times with Horizons},'' \href{http://dx.doi.org/10.1088/0305-4470/10/6/014}{{\em J. Phys. A} {\bfseries 10} (1977) 917--951}.

\bibitem{Akhmedov:2019esv}
E.~T. Akhmedov, K.~V. Bazarov, D.~V. Diakonov, U.~Moschella, F.~K. Popov, and C.~Schubert, ``{Propagators and Gaussian effective actions in various patches of de Sitter space},'' \href{http://dx.doi.org/10.1103/PhysRevD.100.105011}{{\em Phys. Rev. D} {\bfseries 100} no.~10, (2019) 105011}, \href{http://arxiv.org/abs/1905.09344}{{\ttfamily arXiv:1905.09344 [hep-th]}}.

\bibitem{Akhmedov:2020ryq}
E.~T. Akhmedov, P.~A. Anempodistov, K.~V. Bazarov, D.~V. Diakonov, and U.~Moschella, ``{Heating up an environment around black holes and inside de Sitter space},'' \href{http://dx.doi.org/10.1103/PhysRevD.103.025023}{{\em Phys. Rev. D} {\bfseries 103} no.~2, (2021) 025023}, \href{http://arxiv.org/abs/2010.10877}{{\ttfamily arXiv:2010.10877 [hep-th]}}.

\bibitem{Akhmedov:2020qxd}
E.~T. Akhmedov, K.~V. Bazarov, D.~V. Diakonov, and U.~Moschella, ``{Quantum fields in the static de Sitter universe},'' \href{http://dx.doi.org/10.1103/PhysRevD.102.085003}{{\em Phys. Rev. D} {\bfseries 102} no.~8, (2020) 085003}, \href{http://arxiv.org/abs/2005.13952}{{\ttfamily arXiv:2005.13952 [hep-th]}}.

\bibitem{Anempodistov:2020oki}
P.~A. Anempodistov, ``{Remarks on the thermofield double state in 4D black hole background},'' \href{http://dx.doi.org/10.1103/PhysRevD.103.105008}{{\em Phys. Rev. D} {\bfseries 103} no.~10, (2021) 105008}, \href{http://arxiv.org/abs/2012.03305}{{\ttfamily arXiv:2012.03305 [hep-th]}}.

\bibitem{Akhmedov:2021cwh}
E.~T. Akhmedov and D.~V. Diakonov, ``{Free energy and entropy in Rindler and de Sitter space-times},'' \href{http://dx.doi.org/10.1103/PhysRevD.105.105003}{{\em Phys. Rev. D} {\bfseries 105} no.~10, (2022) 105003}, \href{http://arxiv.org/abs/2112.14794}{{\ttfamily arXiv:2112.14794 [hep-th]}}.

\bibitem{Bazarov:2021rrb}
K.~V. Bazarov, ``{Notes on peculiarities of quantum fields in space{\textendash}times with horizons},'' \href{http://dx.doi.org/10.1088/1361-6382/ac8f0e}{{\em Class. Quant. Grav.} {\bfseries 39} no.~21, (2022) 217001}, \href{http://arxiv.org/abs/2112.02188}{{\ttfamily arXiv:2112.02188 [hep-th]}}.

\bibitem{Akhmedov:2022qpu}
E.~T. Akhmedov and K.~V. Bazarov, ``{Backreaction issue for the black hole in de Sitter spacetime},'' \href{http://dx.doi.org/10.1103/PhysRevD.107.105012}{{\em Phys. Rev. D} {\bfseries 107} no.~10, (2023) 105012}, \href{http://arxiv.org/abs/2212.06433}{{\ttfamily arXiv:2212.06433 [hep-th]}}.

\bibitem{Diakonov:2023hzg}
D.~V. Diakonov and K.~V. Bazarov, ``{Debye mass in the accelerating frame},'' \href{http://dx.doi.org/10.1134/S0040577925050095}{{\em Theor. Math. Phys.} {\bfseries 223} no.~2, (2025) 839--862}, \href{http://arxiv.org/abs/2301.07478}{{\ttfamily arXiv:2301.07478 [hep-th]}}.

\bibitem{Kabat:1995eq}
D.~N. Kabat, ``{Black hole entropy and entropy of entanglement},'' \href{http://dx.doi.org/10.1016/0550-3213(95)00443-V}{{\em Nucl. Phys. B} {\bfseries 453} (1995) 281--299}, \href{http://arxiv.org/abs/hep-th/9503016}{{\ttfamily arXiv:hep-th/9503016}}.

\bibitem{Solodukhin:1994yz}
S.~N. Solodukhin, ``{The Conical singularity and quantum corrections to entropy of black hole},'' \href{http://dx.doi.org/10.1103/PhysRevD.51.609}{{\em Phys. Rev. D} {\bfseries 51} (1995) 609--617}, \href{http://arxiv.org/abs/hep-th/9407001}{{\ttfamily arXiv:hep-th/9407001}}.

\bibitem{Fursaev:1994ea}
D.~V. Fursaev and S.~N. Solodukhin, ``{On one loop renormalization of black hole entropy},'' \href{http://dx.doi.org/10.1016/0370-2693(95)01290-7}{{\em Phys. Lett. B} {\bfseries 365} (1996) 51--55}, \href{http://arxiv.org/abs/hep-th/9412020}{{\ttfamily arXiv:hep-th/9412020}}.

\bibitem{Solodukhin:1995ak}
S.~N. Solodukhin, ``{One loop renormalization of black hole entropy due to nonminimally coupled matter},'' \href{http://dx.doi.org/10.1103/PhysRevD.52.7046}{{\em Phys. Rev. D} {\bfseries 52} (1995) 7046--7052}, \href{http://arxiv.org/abs/hep-th/9504022}{{\ttfamily arXiv:hep-th/9504022}}.

\bibitem{Diakonov:2025wtt}
D.~Diakonov, ``{De Sitter entropy: on-shell versus off-shell},'' \href{http://arxiv.org/abs/2504.01942}{{\ttfamily arXiv:2504.01942 [hep-th]}}.

\bibitem{Dodelson:2023vrw}
M.~Dodelson, C.~Iossa, R.~Karlsson, and A.~Zhiboedov, ``{A thermal product formula},'' \href{http://dx.doi.org/10.1007/JHEP01(2024)036}{{\em JHEP} {\bfseries 01} (2024) 036}, \href{http://arxiv.org/abs/2304.12339}{{\ttfamily arXiv:2304.12339 [hep-th]}}.

\bibitem{Grewal:2024emf}
M.~Grewal and Y.~T.~A. Law, ``{Real-time observables in de Sitter thermodynamics},'' \href{http://arxiv.org/abs/2403.06006}{{\ttfamily arXiv:2403.06006 [hep-th]}}.

\bibitem{Lopez-Ortega:2006aal}
A.~Lopez-Ortega, ``{Quasinormal modes of D-dimensional de Sitter spacetime},'' \href{http://dx.doi.org/10.1007/s10714-006-0335-9}{{\em Gen. Rel. Grav.} {\bfseries 38} (2006) 1565--1591}, \href{http://arxiv.org/abs/gr-qc/0605027}{{\ttfamily arXiv:gr-qc/0605027}}.

\bibitem{Anninos:2020hfj}
D.~Anninos, F.~Denef, Y.~T.~A. Law, and Z.~Sun, ``{Quantum de Sitter horizon entropy from quasicanonical bulk, edge, sphere and topological string partition functions},'' \href{http://dx.doi.org/10.1007/JHEP01(2022)088}{{\em JHEP} {\bfseries 01} (2022) 088}, \href{http://arxiv.org/abs/2009.12464}{{\ttfamily arXiv:2009.12464 [hep-th]}}.

\bibitem{Papadodimas:2012aq}
K.~Papadodimas and S.~Raju, ``{An Infalling Observer in AdS/CFT},'' \href{http://dx.doi.org/10.1007/JHEP10(2013)212}{{\em JHEP} {\bfseries 10} (2013) 212}, \href{http://arxiv.org/abs/1211.6767}{{\ttfamily arXiv:1211.6767 [hep-th]}}.

\bibitem{gradshteyn2007}
I.~S. Gradshteyn and I.~M. Ryzhik, {\em Table of integrals, series, and products}.
\newblock Elsevier/Academic Press, Amsterdam, seventh~ed., 2007.
\newblock Translated from the Russian, Translation edited and with a preface by Alan Jeffrey and Daniel Zwillinger, With one CD-ROM (Windows, Macintosh and UNIX).

\bibitem{Birmingham:2001hc}
D.~Birmingham, ``{Choptuik scaling and quasinormal modes in the AdS / CFT correspondence},'' \href{http://dx.doi.org/10.1103/PhysRevD.64.064024}{{\em Phys. Rev. D} {\bfseries 64} (2001) 064024}, \href{http://arxiv.org/abs/hep-th/0101194}{{\ttfamily arXiv:hep-th/0101194}}.

\bibitem{Lin:2022nss}
H.~Lin and L.~Susskind, ``{Infinite Temperature's Not So Hot},'' \href{http://arxiv.org/abs/2206.01083}{{\ttfamily arXiv:2206.01083 [hep-th]}}.

\bibitem{Milekhin:2024vbb}
A.~Milekhin and J.~Xu, ``{On scrambling, tomperature and superdiffusion in de Sitter space},'' \href{http://arxiv.org/abs/2403.13915}{{\ttfamily arXiv:2403.13915 [hep-th]}}.

\end{thebibliography}\endgroup
\bibliographystyle{utphys.bst}
\end{document}